\documentclass[twocolumn,notitlepage,pra,  superscriptaddress]{revtex4-1}
\usepackage{graphicx}
\usepackage{braket}
\usepackage{amsmath,amsfonts,amssymb,amstext,amscd,amsthm,bbold}
\usepackage{comment,float}

\usepackage[colorlinks=true,linkcolor=blue,urlcolor=magenta,citecolor=magenta]{hyperref}
\usepackage[normalem]{ulem}
\begin{document}
\title{Optimal quantum simulation of open quantum systems}
\author{Pragati Gupta}
\email{pragatigupta@iisc.ac.in}
\affiliation{Indian Institute of Science, C.V. Raman Avenue, Bengaluru 560012, India}
\author{C. M. Chandrashekar}
\email{chandru@imsc.res.in}
\affiliation{The Institute of Mathematical Sciences, C. I. T. Campus, Taramani, Chennai 600113, India}
\affiliation{Homi Bhabha National Institute, Training School Complex, Anushakti Nagar, Mumbai 400094, India}
\affiliation{Dept. of Instrumentation \& Applied Physics, Indian Institute of Sciences, C.V. Raman Avenue, Bengaluru 560012, India}

\begin{abstract}
Digital quantum simulation on quantum systems require algorithms that can be implemented using finite quantum resources. Recent studies have demonstrated digital quantum simulation of open quantum systems on Noisy Intermediate-Scale Quantum (NISQ) devices. In this work, we develop quantum circuits for optimal simulation of Markovian and Non-Markovian open quantum systems. The circuits use ancilla qubits to simulate the environment, and memory effects are induced by storing information about the system on extra qubits. We simulate the amplitude damping channel and dephasing channel as examples of the framework and infer (Non-)Markovianity from the (non-)monotonic behaviour of the dynamics. Further, we develop a method to optimize simulations by decomposing complex open quantum dynamics into smaller parts, that can be simulated using a small number of qubits.  We show that this optimization reduces quantum space complexity from $O(l)$ to $O(1)$ for simulating the environment. 
\end{abstract}
\maketitle

%========================================
\section{Introduction}
%========================================

Real world quantum systems interact with the environment as opposed to the idealized notion of closed quantum systems. The theory of open quantum systems\,\cite{breuer2002theory,rivas2012open} provides a useful framework to study a variety of phenomena, such as the decoherence in quantum systems\,\cite{opendecoherence}, out of equilibrium many-body dynamics\,\cite{manybody}, quantum field theory\,\cite{qft}  and has applications to studying information processing in the physical systems. A description of open quantum systems in the form of partial differential equation of motion is provided by the quantum master equations. These equations are derived under certain assumptions about the dynamics and a general form of the master equation for an open quantum system is not known. For example, the Lindblad master equation\,\cite{lindblad}, holds under the Born-Markov approximation and can not be used to describe Non-Markovian systems. Another formulation, the Kraus representation\,\cite{book} provides a more general framework and has been widely used to characterize open quantum systems. Quantum simulation of open quantum systems can provide insights into system-environment interactions, that can be used to understand and control noisy physical systems\,\cite{johansson2012qutip}. \\

Quantum simulators are devices that can turn the exponential scaling of resources needed to simulate quantum systems using classical systems into a favourable polynomial overhead. Ongoing experimental research is geared towards making quantum simulators with small number of qubits, called Noisy Intermediate Scale Quantum (NISQ) devices as first practical applications of a quantum computer\,\cite{nisq,nearterm}. Quantum simulation of open quantum systems usually relies on using a simulator that directly mimics the dynamics of the quantum system under interest. This approach is useful in near term analog quantum simulations and has been applied to a number of systems\,\cite{analog2,analog3,analog4,mostame2017emulation,mostame2012quantum}. In contrast, a universal quantum computer\,\cite{deutsch1985quantum} or a quantum Turing machine\,\cite{bernstein1997quantum} provides a model to harness the complete power of programmable quantum computation. Digital quantum simulation is a gate-based approach where single and two-qubit quantum gates are used to implement quantum operations and is closely related to universal quantum computation\,\cite{arrazola2016digital}. Quantum circuits are commonly used models for digital quantum simulation\,\cite{yao1993quantum}, where states in a Hilbert space are acted upon by quantum operations using a collection of unitary matrices that map the Hilbert space to itself\,\cite{nielsen1997programmable}. Digital quantum simulation of open quantum systems has gained attention only recently\,\cite{digital1,digital2,digital3,perez,digital4,digital5,digital6,2,3,4,5,6,nathan2020universal,muller2011simulating,sweke2014simulation,naikoo2020nonmarkovian}. In \,\cite{perez}, IBM Q-Experience simulator has been used to implement paradigmatic open quantum dynamics such as amplitude damping and entanglement pair generation. Discrete time quantum algorithms for studying quantum channels has also been explored\,\cite{2,3,4,5,6,naikoo2020nonmarkovian}.\\

The limited resources on a NISQ device poses several challenges for digital quantum simulation\,\cite{preskill2020quantum}. In this work, we present a general methodology to design optimal quantum circuits for Markovian and Non-Markovian open quantum systems. The methodology is developed systematically: First, we use the notion of P-divisibility\,\cite{divisibility1,divisibility2,divisibility3} of a quantum channel to formulate operators for Markovian dynamics. We use ancilla qubits to simulate the environment and note that tracing out the ancilla induces memory-less evolution. Then, we propose a scheme for Non-Markovian dynamics by duplicating the information on extra qubits, which are not traced out and thus retain memory. We use a discrete-time approach and simulate memory effects up to a finite order (i.e., the contributions from a finite number of states in the past). Using the ancilla, we create a quantum "register" to store information about past states, and update the register at the end of every step using SWAP gates. We demonstrate simulation of amplitude damping and dephasing of a qubit as examples. Non-Markovianity, in the sense of P-indivisibility of the quantum circuits is inferred from the non-monotonous behavior of the dynamics. These examples show that the methodology is useful for simple open quantum system. Then, we generalize it to more complex dynamics. We show that complex open quantum dynamics can be decomposed into smaller processes with finite resource requirements. The complete process, which generally needs parallel processing can be implemented sequentially using the decomposition. Converting the parallel process into a sequential one  optimizes quantum space requirements, thus less number of qubits are required. Since gate decomposition into elementary operations becomes easier, number of gates also gets optimized. We simulate a Pauli channel using this method, and show a reduction in number of qubits compared to known methods.\\ 

This article is organized as follows. In Sec.\,\ref{sec:oqs}, an overview of open quantum dynamics is given. In Sec.\,\ref{sec:quantumcircuits} quantum circuit for implementing the Markovian and Non-Markovian dynamics are derived and in Sec.\,\ref{sec:simulations} numerical simulations of the quantum circuit models are presented is presented. In sec.\,\ref{sec:optimization}, a methodology for optimizing the quantum circuits is presented and the quantum complexity analysis of implementation is discussed. Potential applications of this work are discussed in Sec.\,\ref{sec:conclusion}.
%~\newpage~\newpage

%===============================================================================================================================
\section{Open quantum systems}\label{sec:oqs}
The state of a quantum mechanical system is denoted by its wavefunction $\ket{\psi}$ and the evolution of a system is given by the Schrodinger equation,
\begin{equation}\label{eq:schrodinger}
    i\hbar\frac{d}{dt}\ket{\psi(t)}=H\ket{\psi(t)}.
\end{equation}
where $H$ is the Hamiltonian operator of the system. The Hamiltonian captures the sum of kinetic and potential energy of the system. The solution of the Sch\"{o}dinger equation is given by,
\begin{equation}\label{eq:unitary}
    \ket{\psi(t)}=U\ket{\psi(0)}=e^{-i\frac{H}{\hbar}t}\ket{\psi(0)}.
\end{equation}
The evolution is unitary as described by the operator $U$. In terms of the density matrix $\rho$, the Schr\"{o}dinger equation is written as,
\begin{equation}\label{eq:rhoschrodinger}
    \frac{d\rho}{dt}=-i[H,\rho].
\end{equation}
Solving the above equation gives,
\begin{equation}\label{eq:rhounitary}
	\rho(t)=U\rho(0)U^\dagger.
\end{equation}
The above holds true for closed quantum systems. An open quantum system is a part of a larger closed system, composed of the system $S$ and its environment $E$. The Hilbert space of the total system $S+E$ is given by
\begin{equation} \label{Hilbert-total}
 {\mathcal{H}}_{SE} = {\mathcal{H}}_S\otimes{\mathcal{H}}_E,
\end{equation}
where ${\mathcal{H}}_S$ and ${\mathcal{H}}_E$ denote the Hilbert spaces of $S$ and $E$, respectively. \\ 

To describe the evolution of an open quantum system, one considers the evolution of the total system and traces out the environment at the end. A common assumption made is that the initial state of the total system is of the form,
\begin{equation} \label{Product-initial-state}
 \rho_{SE}(0) = \rho_S(0) \otimes \rho_E(0).
\end{equation}
The state of the system at any time $t\geq 0$ is then described by,
\begin{equation} \label{Rho-S-repr}
 \rho_S(t) = {\rm{tr}}_E \left\{ U(t) \rho_S(0) \otimes \rho_E(0) U^{\dagger}(t) 
 \right\}.
\end{equation}
This maps any initial state $\rho_S(0)$ to its final state using a dynamical map $\Phi_t: \, S({\mathcal{H}}_S) \longrightarrow S({\mathcal{H}}_S)$\,\cite{divisibility2}, such that,
\begin{equation} \label{Dynmap}
 \rho_S(0) \mapsto  \rho_S(t) = \Phi_t \rho_S(0).
\end{equation}
$\Phi_t$  is a positive map and thus, maps physical states to physical states. Further, $\Phi_t$ is completely positive and admits a Kraus operator representation,
\begin{equation}\label{kraus}
    \Phi A  = \sum_i \Omega_i A \Omega_i^{\dagger},
\end{equation}
for operator $A$, where $\Omega_i$ are Kraus operators and $\sum_i \Omega_i^{\dagger}\Omega_i = I_S$.
Open quantum dynamics can be Markovian or Non-Markovian. This characterization is closely related to the divisibility of the dynamical map. We describe these concepts below. %That is, the flow of information from the system(S) to the environment(E) 

\subsection{Divisibility and Markovianity}
Let us define $\Phi_{t,s}$ as,
\begin{equation} \label{two-parameter-family}
 \Phi_{t,s} = \Phi_t \Phi_s^{-1}, \qquad t\geq s \geq 0,
\end{equation}
assuming that the inverse of $\Phi_t$ exists for all times $t\geq 0$, such that,
\begin{equation} \label{divisibility1}
 \Phi_{t,0}=\Phi_t ~~~~\mbox{and}~~~~~ \Phi_{t,0} = \Phi_{t,s} \Phi_{s,0}.
\end{equation}
Even though $\Phi_{t,0}$ and $\Phi_{s,0}$ are completely positive maps, the map $\Phi_{t,s}$ need not be completely positive and not even positive since the inverse $\Phi^{-1}_s$ of a completely positive map $\Phi_s$ need not be positive. \\ 

Eq.\eqref{divisibility1} allows us to introduce the notion of divisibility. The family of dynamical maps $\Phi = \left\{ \Phi_t \mid 0 \leq t \leq T, \Phi_0 = I \right\}$ is said to be P-divisible if $\Phi_{t,s}$ is positive, and CP-divisible if $\Phi_{t,s}$ is completely positive for all $t\geq s \geq 0$. \\ 

Divisibility of a dynamical map is closely related to the Markovianity of evolution. In a Markovian process, the state of the system at the next moment depends only on the current state of the system. In other words, the system doesn't have a memory of its previous states. In contrast,  Non-Markovian evolution depends on the previous states of the system. That is, the system has memory of its previous states. In general, Markovian processes are P-divisible, while Non-Markovian processes are not P-divisible. As we will see in the next section, this classification allows us to derive models for digital quantum simulation of open quantum systems.

%============================================================================================================================

\section{Quantum circuits for open systems}\label{sec:quantumcircuits}
%==================================================================

In this section, we give an overview of Markovian and Non-Markovian dynamics, and describe the quantum circuits for simulating open quantum systems.
%==================================
\subsubsection{Markovian dynamics}
%==================================
To simulate open quantum systems on a quantum circuit, we can use ancillary qubits to mimic the effect of environment. As discussed in the previous section, Markovian dynamics have no memory of the previous states of the system. That is, state at $n^{th}$ step depends only on the ${n-1}^{th}$ step. Since the evolution only depends on the current state of the system, the environment need not store information and the ancilla can be reset after each step.  \\

For Markovian evolution, $\rho_S(t+\delta t)$ only depends on $\rho_S(t)$. Thus, in the discrete time setup, the evolution can be described as,
\begin{equation}\label{Markovian-Kraus}
    \rho_S(t+\Delta t)= \sum_{i} \Omega_i \rho_S(t) \Omega_i^{\dagger} = \Phi_{\Delta t}\rho_S(t). 
\end{equation}
The second equality holds due to divisibility of Markovian dynamics (\,\,Eq.\ref{divisibility1}). \,\,Eq.\eqref{Markovian-Kraus} can be represented as a unitary evolution over system(S) + environment(E) such that,
\begin{equation}\label{markovian-unitary}
    \begin{split}
        \rho_{SE}(t)=&\rho_S(t)\otimes{\ket{0}\bra{0}}_E, \text{and},\\
        \rho_{SE}(t+\Delta t)=&\sum_i\Omega_i \rho_S(t) \Omega_i^{\dagger}\otimes{\ket{i}\bra{i}}_E,
    \end{split}
\end{equation}
where $\ket{i}_E$ are the basis states of the environment. Tracing out the environment from the above equation gives back \,\,Eq.\eqref{Markovian-Kraus}. The unitary evolution operator can be written as (using Stinespring dilation)\,\cite{preskillnotes},
\begin{equation}\label{markovian-operator}
    \begin{split}
        &U_{SE}(\Delta t)=\sum_i\Omega_i\otimes{\ket{i}\bra{0}}_E\\
        &\rho_{SE}(t+\Delta t)=U_{SE} \Big(\rho_S(t) \otimes{\ket{0}\bra{0}}_E\Big) U_{SE}^{\dagger}
    \end{split}
\end{equation}
It can be verified that $U_{SE}^\dagger U_{SE}=I_S\otimes \ket{0}\bra{0}_E$. This can be used to simulate Markovian evolution by tracing out environment at the end of each step and resetting it to $\ket{0}$ at the beginning of next step.

%==================================
\subsubsection{Non-Markovian dynamic}
%==================================
In \,\,Eq.\eqref{markovian-operator}, when we trace out the environment, we discard any information about the previous step from the ancilla. This leads to memory less evolution. The key idea behind simulating Non-Markovian systems is to retain some information about the system in the ancillas. \\

Take an environment consisting of two qubits $E_1$ and $E_2$. Let $\ket{i,j}_E=\ket{i}_{E_1}\ket{j}_{E_2}$, where $\ket{i}_E$ are the basis states of the environment. Let,
\begin{equation}\label{nonmarkovian-unitary}
    %\begin{split}
        %\rho_{SE}(t)=&\rho_S(t)\otimes{\ket{0}\bra{0}}_E\\
        \rho_{SE}(t+\Delta t)=\sum_i \Omega_i \rho_S(t) \Omega_i^{\dagger}\otimes{\ket{i,i}\bra{i,i}}_E.
    %\end{split}
\end{equation}
Here, we have duplicated the information on two environment qubits instead of one qubit. So, if we trace out one of these, we will still have information about this step on the other qubit. 

After tracing out $E_2$ from \,\,Eq.\eqref{nonmarkovian-unitary} and resetting it to $\ket{0}_{E_2}$, we get
\begin{equation}
    %\begin{split}
        \rho_{SE}(t+\Delta t)=\sum_i \Omega_i \rho_S(t) \Omega_i^{\dagger}\otimes{\ket{i,0}\bra{i,0}}_E.
    %\end{split}
\end{equation}
Now, if qubit $E_1$ controls the dynamics in the next step, this will lead to backflow of information. Thus, we can create a memory for the system using this method, leading to Non-Markovian dynamics. \\

In principle, we would need an infinite system to store information about the state at all times in the past. However, we can overcome this by considering a discrete-time setup (also suited for digital quantum simulation) and neglect contributions beyond a certain order. That is, evolution at $n^{th}$ step depends on last $k$ steps for some finite $k>1$. In such a case, we only need to store information about $(n-1)^{th}$, $(n-2)^{th}$,\ldots, $(n-(k-1))^{th}$ steps. Here, we are taking $k$ orders of contribution to evolution including the contribution from the current ($n^{th}$) state. Thus, we need a way to "store" the information about last $k-1$ steps, and update this information after each new step. In addition we need to discard some information at the end of each step. For example, for $k=3$, for $n^{th}$ step, we need to store information about $(n-1)^{th}$ and $(n-2)^{th}$ step. Then, for $n+1)^{th}$ step, we need to store information about $(n)^{th}$ and $(n-1)^{th}$ step, thus discarding information about $(n-2)^{th}$ step, and so on. This way, we only need to store information on $k$ qubits at a time. Thus, we need qubits $E_1,\ldots,E_k$ for the environment.\\

Below, we describe in detail the quantum circuits for simulating Markovian and Non-Markovian open quantum systems. We illustrate the methodology with the help of two examples, amplitude damping and phase damping on a qubit. Later, we show that this can be extrapolated to any open quantum system. 

\subsection{Example 1: Amplitude damping}

\subsubsection{Markovian}
Amplitude damping of a qubit is represented by Kraus operators
\begin{equation}
   \Omega_0=\ket{0}\bra{0}+\sqrt{1-\gamma^2}\ket{1}\bra{1}, \qquad  \Omega_1=\gamma\ket{0}\bra{1}.
\end{equation}
The unitary operator for Markovian evolution ($U_{SE}(\Delta t)= \Omega_0\otimes{\ket{0}\bra{0}}_E+ \Omega_1\otimes{\ket{1}\bra{0}}_E$) can be written as,
\begin{equation}\label{amplitudeDTQW}
\begin{split}
    &U_{SE}=\hat{S}\hat{C}, \text{where}\\
    &\hat{C}=\ket{0}\bra{0}_S\otimes\ket{0}\bra{0}_E + \ket{1}\bra{1}_S\otimes R_y(\theta)\ket{0}\bra{0}_E, \\
    &\hat{S}=I_S\otimes\ket{0}\bra{0}_E + NOT_S\otimes \ket{1}\bra{1}_E.
\end{split}
\end{equation}
Here, $\sin{\frac{\theta}{2}}=\gamma$  and $R_y$ is the rotation matrix about y-axis on the Bloch sphere: $R_y(\theta)\ket{0}=\sqrt{1-\gamma^2}\ket{0}+\gamma\ket{1}$. To verify \,\,Eq.\eqref{amplitudeDTQW},
\begin{equation}
    \begin{split}
    \hat{S}\hat{C}=&\hat{S}.\{\ket{0}\bra{0}_S\otimes\ket{0}\bra{0}_E \\
    &+ \ket{1}\bra{1}_S\otimes \sqrt{1-\gamma^2}\ket{0}\bra{0}_E\\
    &+ \ket{1}\bra{1}_S\otimes \gamma\ket{1}\bra{0}_E\}\\
    =&\ket{0}\bra{0}_S\otimes\ket{0}\bra{0}_E\\
    &+\sqrt{1-\gamma^2}\ket{1}\bra{1}_S\otimes \ket{0}\bra{0}_E\\
    &+\gamma\ket{0}\bra{1}_S\otimes \ket{1}\bra{0}_E\\
    =&\{\Omega_0\otimes{\ket{0}\bra{0}}_E+ \Omega_1\otimes{\ket{1}\bra{0}}_E\}.
    \end{split}
\end{equation}
This decomposition of the unitary evolution helps us simulate amplitude damping using elementary gates. \\
\begin{figure}
    \centering
    \includegraphics{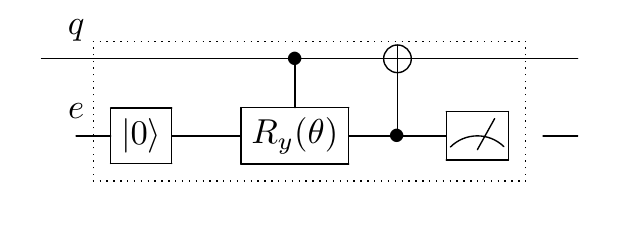}
    \caption{Quantum circuit for implementing one step of Markovian amplitude damping of a qubit. Here, $q$ and $e$ represent the qubit and environment respectively. $e$ is set to $\ket{0}$ at the beginning of every step. Then, the rotation operation on $e$ sets the right control to implement amplitude damping using a $CNOT$ gate. The environment is traced out at the end of each step. We can sequentially implement these steps to capture the complete evolution.}
    \label{fig:Markovian amplitude damping}
\end{figure}

Fig.\,\ref{fig:Markovian amplitude damping} shows the quantum circuit for implementing amplitude damping of a qubit. The unitary operator $U_{SE}$ is implemented using $\hat{C}$ and $\hat{S}$ operators. $\hat{C}$ is implemented using the controlled rotation gate and $\hat{S}$ operator is implemented using the $CNOT$ gate. The environment is traced out at the end of a step to induce Markovianity in the dynamics. The figure shows one step of the dynamics (comprising of operations in the box), which can be repeated to implement the complete evolution. \\

Thus, by mapping an amplitude damping channel to unitary evolution on a larger system, we could design quantum circuits for its simulation. Here, we considered a Markovian process. Next, we formulate the quantum circuits for Non-Markovian amplitude damping.

\subsubsection{Non-Markovian}
In the above case, the $R_y(\theta)$ gate is used to control the effect of the environment on the qubit. $\theta$ captures the first order contribution to the Markovian dynamics. When we trace out $e$ in Fig.\,\ref{fig:Markovian amplitude damping}, we discard all the information about the previous state of the system. To introduce memory effects, we use extra environment qubits which are able to store this information and simulate higher order contributions in Non-Markovian systems. Below, we illustrate and explain the quantum circuit implementation for Non-Markovian amplitude damping of a qubit, with three orders of contributions (k=3) to memory in the dynamics.\\
\begin{figure}
    \centering
    \includegraphics[width=0.5\textwidth]{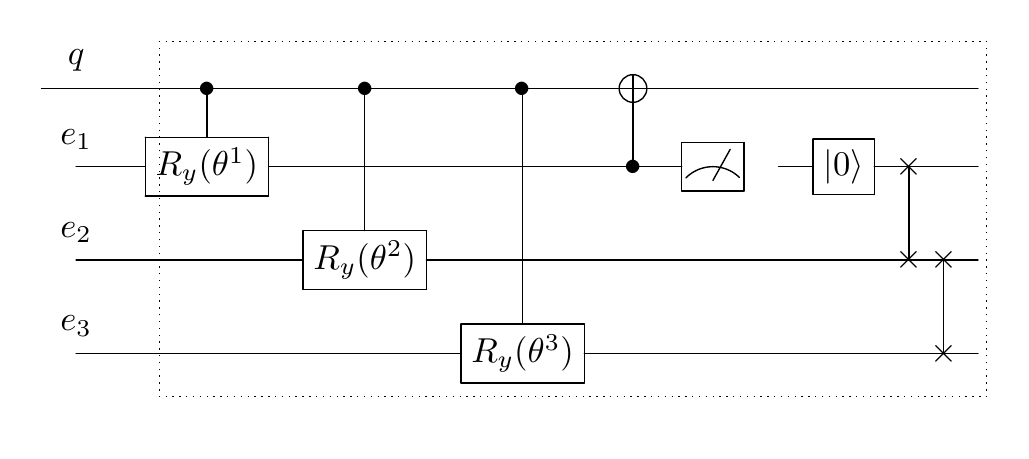}
    \caption{Quantum circuit for implementing one step of Non-Markovian amplitude damping of a qubit. Here, $q$ and $e_i$ represent the qubit and environment respectively. $R_y(\theta_i)$ is used to "store" contributions of current state to $i^{th}$ order dynamics.  Amplitude damping of the qubit is implemented using a $CNOT$ gate. $e_1$ is traced out at the end of each step and reset to $\ket{0}$ and SWAP gates are used to update the environment states before the next step. We can sequentially implement these steps to capture the complete evolution (Fig.\,\ref{fig:4steps}).}
    \label{fig:1step}
\end{figure}
Fig.\,\ref{fig:1step} shows the quantum circuit for implementing one step of Non-Markovian evolution. Here, $q$ represents the qubit under amplitude damping and $e_1, e_2, e_3$ represent the environment. Controlled rotation matrices $R_y(\theta^i)$, for $i\in{1,2,3}$ are used to capture the contribution of the current state to $i^{th}$ order dynamics. $i^{th}$ order dynamics at a particular time $t$, is due to contribution from state at time $t-i$. Thus, we need to store information about a state for future retrieval. The controlled-$R_y(\theta^1)$ operation on $e_1$ captures first order contribution, similar to the Markovian case (Fig.\,\ref{fig:Markovian amplitude damping}). Then, we store information about the current state on $e_2$ and $e_3$ for contributions to future steps using controlled-$R_y(\theta^{2/3})$. The $CNOT$ gate is used to implement the amplitude damping ($\ket{1}\to\ket{0}$) in current state. $e_1$ is then traced out and reset to $\ket{0}$ state.\\

After this, the first $SWAP$ gate exchanges the state of $e_1$ and $e_2$. Thus, the state of $e_1$ becomes $R_y(\theta^2)\ket{0}$ and the state of $e_2$ becomes $\ket{0}$.  Similarly, after the next SWAP gate, the state of $e_2$ becomes $R_y(\theta^3)\ket{0}$ and the state of $e_3$ becomes $\ket{0}$. Here, we have assumed for now that $e_2$ and $e_3$ were initially in state $\ket{0}$. So, the states of ${e_1}, {e_2}, \text{ and } {e_3}$ are $R_y(\theta^2)\ket{0}, R_y(\theta^3)\ket{0}, \text{ and } \ket{0}$ respectively. Thus, before the next step $e_1$ is is correlated with the qubit and carries information about the previous step. \\

\begin{widetext}
\begin{figure}[H]
    \centering
    \includegraphics[width=\textwidth]{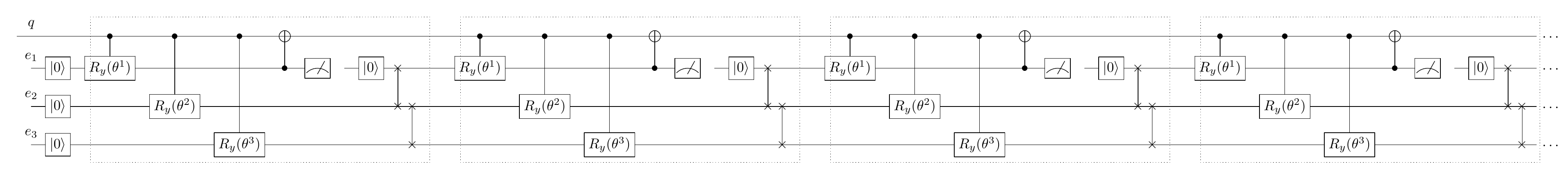}
    \caption{Quantum circuit for Non-Markovian amplitude damping of a qubit, by sequentially implementing the steps (boxes).}
    \label{fig:4steps}
\end{figure}
\end{widetext}

Fig.\,\ref{fig:4steps} shows implementation of complete Non-Markovian evolution by sequential application of steps. In the second step, $e_1$ already carries information from the previous step. Thus, when we implement controlled rotation matrix $R_y(\theta^1)$, $e_1$ carries contributions from both the current ($\theta^1$) and the previous step ($\theta^2$). Similarly, after controlled rotation matrix $R_y(\theta^2)$, $e_2$ carries contributions of $\theta^2$ due to current step and $\theta^3$ due to previous step. After controlled rotation matrix $R_y(\theta^3)$, $e_3$ only carries information due to current step, as it was in state $\ket{0}$ at the beginning. Then, implementing the $CNOT$ gate using $e_1$ in the second step not only captures first order effects due to $\theta^1$, but also second order effects coming from $\theta^2$ which was entangled with the state of the system in the previous step. This capture the memory effect of the bath when we trace out $e_1$ and reset to $\ket{0}$. \\
Applying the two SWAP gates again, leads to $e_1$ carrying contributions due to $\theta^2$ and $\theta^3$, $e_2$ carrying contributions due to $\theta^3$, and $e_3$ is in state $\ket{0}$. Thus, when we implement controlled rotation matrices $R_y(\theta^1)$ of step 3, $e_1$ carries contributions from $\theta^1$ due to current step, and $\theta^2$ and $\theta^3$ due to previous steps. Again, $e_2$ carries contributions of $\theta^2$ due to current step and $\theta^3$ due to previous step. Then, implementing the $CNOT$ gate using $e_1$ in the third step captures first, second and third order contributions. Thus, we are able to model memory effects in the evolution. Similar analysis holds for the following steps, where we $e_1$ always captures contribution from $k=3$ orders of dynamics. \\

Overall, the idea is to trace out only a part of the environment (only $e_1$ is traced out, $e_2$ and $e_3$ are not measured) and then introduce appropriate evolution within the environment to retain some of the coherence for the next step, thus modeling memory effects. In the above example, we used $e_2$ and $e_3$ to store appropriate information about the history of the system, and used $e_1$ to implement the current step. The information was updated at the end of each step using SWAP gates which "shift" the information up by one order to prepare for the next step. This methodology can be extended to any Non-Markovian open quantum system. Next, we use it to model dephasing of a qubit.

\subsection{Example 2: Dephasing of a qubit}

Here, we illustrate another example of Markovian and Non-Markovian process. We follow the same methodology as presented above: First, we write the Kraus representation of the dynamics and use Stinespring dilation to find a unitary operator for simulating the system along with the environment. Then, we decompose the unitary operator into elementary gates to formulate quantum circuits for the dynamics. After obtaining a model for Markovian process, we generalize it to Non-Markovian process by adding extra environment qubits that store information. We partially trace out the environment at the end of a step, and use SWAP gates to update the information stored on qubits for the next step. This helps to model memory effects in the dynamics. Below, we apply this methodology to dephasing of a qubit.\\

Markovian dephasing can be written in the Kraus representation as,
\begin{equation}
\begin{split}
    \rho_S(t+\Delta t)=\Omega_0\rho_S(t)\Omega_0^\dagger +\Omega_1\rho_S(t)\Omega_1^\dagger\\
    \Omega_0=\sqrt{1-\gamma^2}I_S, \qquad \Omega_1=\gamma Z_S.
\end{split}
\end{equation}
The unitary operator for Markovian evolution ($U_{SE}(\Delta t)= \Omega_0\otimes{\ket{0}\bra{0}}_E+ \Omega_1\otimes{\ket{1}\bra{0}}_E$) can be written as,
\begin{equation}\label{phasedampingDTQW}
\begin{split}
    &U_{SE}=\hat{S}\hat{C}\\
    &\hat{C}=I_S\otimes R_y(\theta)\ket{0}\bra{0}_E \\
    &\hat{S}=I_S\otimes\ket{0}\bra{0}_E + Z_S\otimes \ket{1}\bra{1}_E 
\end{split}
\end{equation}
where $\sin{\frac{\theta}{2}}=\gamma$, $R_y(\theta)\ket{0}=\sqrt{1-\gamma^2}\ket{0}+\gamma\ket{1}$. This can be proven as,
\begin{equation}
    \begin{split}
    \hat{S}\hat{C}=\hat{S}.\{I_S\otimes \sqrt{1-\gamma^2}\ket{0}\bra{0}_E+ I_S\otimes \gamma\ket{1}\bra{0}_E&\}\\
    =\sqrt{1-\gamma^2}I_S\otimes\ket{0}\bra{0}_E+\gamma Z_S\otimes \ket{1}\bra{0}_E&\\
    =\{\Omega_0\otimes{\ket{0}\bra{0}}_E+ \Omega_1\otimes{\ket{1}\bra{0}}_E\}&.
    \end{split}
\end{equation}
\begin{figure}
    \centering
    \includegraphics{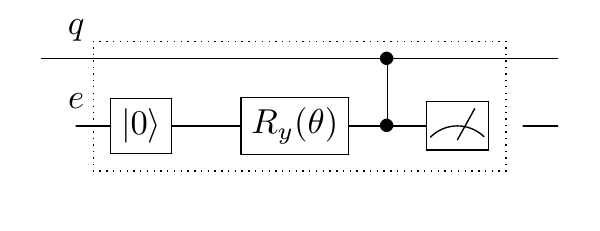}
    \caption{Quantum circuit for implementing one step of Markovian dephasing of a qubit. Here, $q$ and $e$ represent the qubit and environment respectively. $e$ is set to $\ket{0}$ at the beginning of every step. Then, the rotation operation on $e$ sets the right control to implement dephasing using a $CZ$ gate. The environment is traced out at the end of each step. We can sequentially implement these steps to capture the complete evolution.}
    \label{fig:Markovian dephasing}
\end{figure}
Fig.\,\ref{fig:Markovian dephasing} shows the quantum circuit for implementing dephasing of a qubit. The unitary operator $U_{SE}$ is implemented using $\hat{C}$ and $\hat{S}$ operators. $\hat{C}$ is implemented using the rotation gate and $\hat{S}$ operator is implemented using the $CZ$ gate, as shown in the figure. The environment is traced out at the end of a step. One complete step (comprising of operations in the box) can be repeated to implement the complete evolution. \\

\begin{figure}
    \centering
    \includegraphics{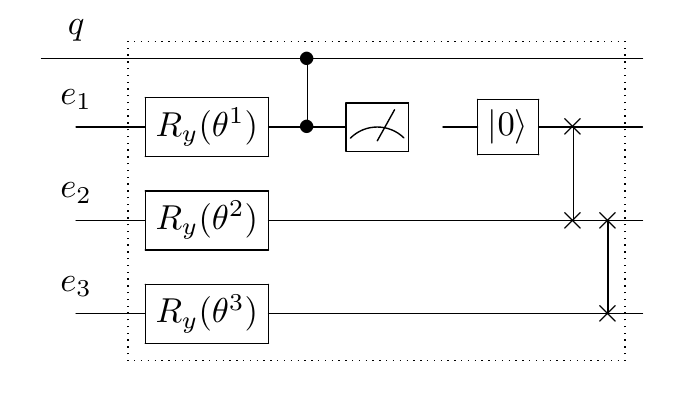}
    \caption{Quantum circuit for implementing one step of Non-Markovian dephasing of a qubit. Here, $q$ and $e_i$ represent the qubit and environment respectively. $R_y(\theta_i)$ is used to "store" contributions of current state to $i^{th}$ order dynamics.  Dephasing of the qubit is implemented using a $CZ$ gate. $e_1$ is traced out at the end of each step and reset to $\ket{0}$ and SWAP gates are used to update the environment states before the next step. We can sequentially implement these steps to capture the complete evolution.}
    \label{fig:NonMarkovian dephasing}
\end{figure}
Fig.\,\ref{fig:NonMarkovian dephasing} shows quantum circuit for dephasing of a qubit with $k=3$ orders of contribution. Similar to the analysis in the previous example, the circuit here will exhibit memory effects and lead to Non-Markovian dynamics.\\

\subsection{Numerical Simulations}\label{sec:simulations}
\begin{figure}
    \centering
    \includegraphics[width=0.5\textwidth]{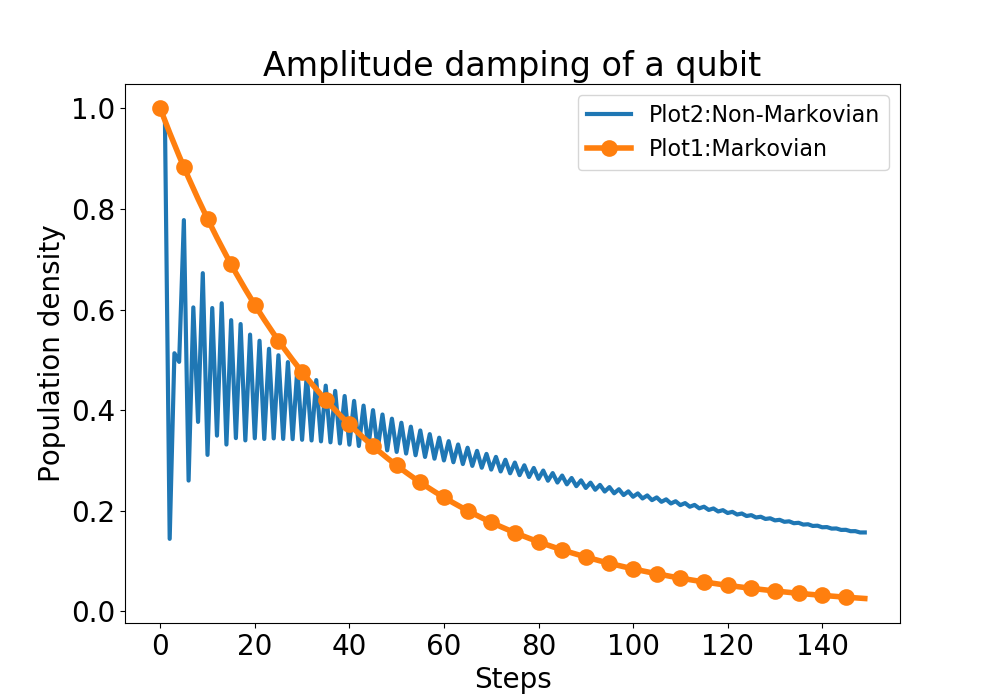}
    \caption{Evolution of the population density of state $\ket{1}$ for a qubit under amplitude damping. Markovian dynamics are calculated according to Fig.\,\ref{fig:Markovian amplitude damping} and Non-Markovian dynamics are calculated according to Fig.\,\ref{fig:1step}.}
    \label{fig:amplitude damping plot}
\end{figure}

Here we present numerical simulations of the quantum circuit models presented above. Fig.\,\ref{fig:amplitude damping plot} shows the evolution of population densities in state $\ket{1}$ of a qubit under Markovian and Non-Markovian amplitude damping. The initial state of the qubit $\rho(0)=\ket{1}\bra{1}$. The dynamics are simulated for the circuits shown in Fig.\,\ref{fig:Markovian amplitude damping} and Fig.\,\ref{fig:1step}. For the Markovian case, the population in $\ket{1}$ slowly dies off and gets transferred to state $\ket{0}$ through amplitude damping. Here, we have taken $\theta=\pi/10$, \,Eq.(\ref{amplitudeDTQW}). For the Non-Markovian case, the population in $\ket{1}$ oscillates while decreasing. Here, we have taken $\theta^1=\pi/10$, $\theta^2=2\pi/3$, and $\theta^3=5\pi/6$, Fig.\,\ref{fig:1step}. From the two plots, we can note that the quantum circuit models capture the features of Markovian and Non-Markovian systems. For plot 1, the decay in population is strictly monotonic, which is characteristic of Markovian evolution. In Non-Markovian evolution, we can observe the non-monotonic behavior due to memory effects of the environment. Comparing the two, we can see that population in state $\ket{1}$ sustains for longer due to memory in Non-Markovian case. 
\begin{figure}
    \centering
    \includegraphics[width=0.5\textwidth]{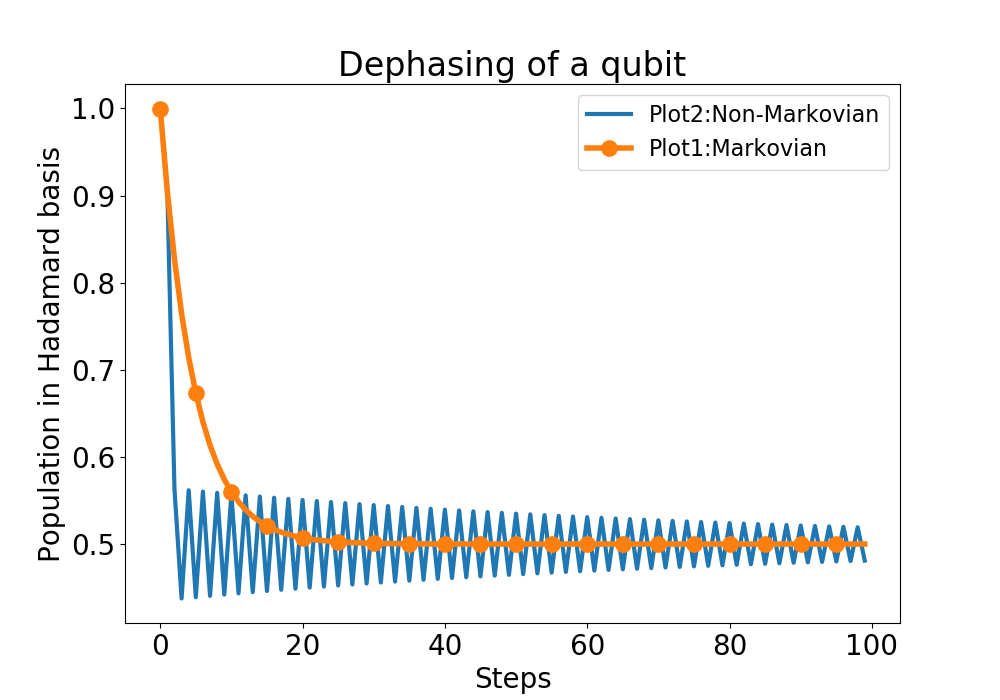}
    \caption{Evolution of the population density of state $\ket{+}$ (Hadamard basis) for a qubit under phase damping. Markovian dynamics are calculated according to Fig.\,\ref{fig:Markovian dephasing} and Non-Markovian dynamics are calculated according to Fig.\,\ref{fig:NonMarkovian dephasing}.}
    \label{fig:dephasing}
\end{figure}

Fig.\,\ref{fig:dephasing} shows the simulations of the phase damping model. The population is plotted in the Hadamard basis to show the effect of phase damping. The initial state of the qubit is $\ket{+}=\frac{\ket{0}+\ket{1}}{\sqrt{2}}$. For the Markovian case, the population in $\ket{+}$ slowly dies off and gets transferred to state $\ket{-}$ through phase damping. Here, we have taken $\theta=\pi/5$, \,Eq.(\ref{phasedampingDTQW}). For the Non-Markovian case, the population in $\ket{+}$ oscillates while decreasing. Here, we have taken $\theta^1=\pi/5$, $\theta^2=\pi/4$, and $\theta^3=\pi/2$, Fig.\,\ref{fig:NonMarkovian dephasing}. For plot 1, the decay in population is strictly monotonic, which is characteristic of Markovian evolution. In Non-Markovian evolution, we can observe the non-monotonic behavior due to memory effects of the environment. Comparing the two, we can see that population in state $\ket{+}$ oscillates for longer due to memory in Non-Markovian case. In both plots, the final population tends to $0.5$. 
\begin{figure}
    \centering
    \includegraphics[width=0.5\textwidth]{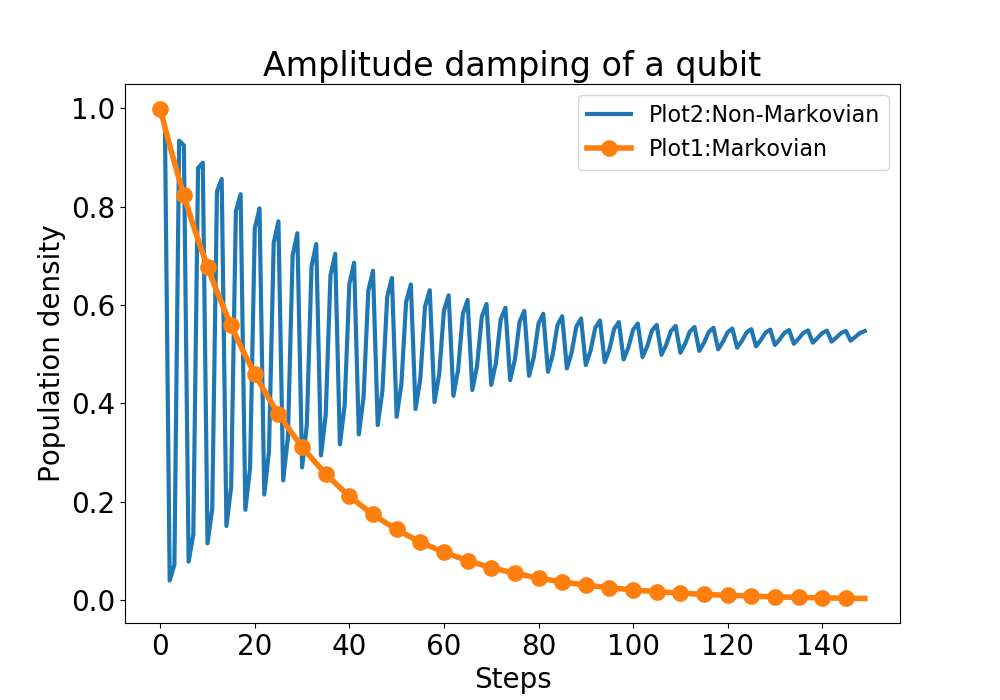}
    \caption{Evolution of the population density of state $\ket{1}$ for a qubit under amplitude damping. Due to memory effects, population in state $\ket{1}$ in Non-Markovian dynamics sustains despite amplitude damping. Markovian dynamics are calculated according to Fig.,\ref{fig:Markovian amplitude damping} and Non-Markovian dynamics are calculated according to Fig.\,\ref{fig:1step}.}
    \label{fig:interesting}
\end{figure}

Fig.\,\ref{fig:interesting} shows another example of amplitude damping of a qubit. Here, we take $\theta^1=\pi/8$ for the Markovian case, and $\theta^1=\pi/8$, $\theta^2=5\pi/6$, and $\theta^3=\pi$ for the Non-Markovian case. We can note that this figure greatly differs from Fig.\,\ref{fig:amplitude damping plot}. Here, the memory effects lead to population in state $\ket{1}$ to sustain despite amplitude damping. This shows that Non-Markovian phenomena are quite interesting and can show varied dynamics.\\

Thus, the quantum circuit models presented in the two examples capture the appropriate Markovian and Non-Markovian dynamics. These examples can be used to extrapolate the methodology to other cases. In the next section, we analyze the complexity of the quantum circuits and optimize the implementation for general open quantum systems,

%==================================================================
\section{Complexity analysis and optimization}\label{sec:optimization}
In amplitude damping (or phase damping) through a Markovian process, we used $1$ qubit to represent the system, and one qubit to simulate the environment, Fig.\,\ref{fig:Markovian amplitude damping}, and \ref{fig:Markovian dephasing}. In general, for a system with Hilbert space dimension $n$, we need $\log n$ qubits for simulation. We had arrived at a quantum circuit model for open quantum systems using Stinespring dilation of the Kraus representation, ,Eq.(\ref{markovian-unitary}). For a Kraus representation consisting of $l$ terms in the sum, we need atleast $\log l$ qubits for the environment. For the amplitude damping (or phase damping) case, $l=2$, thus one qubit was needed for simulating the environment. In general, $O(l)$ different qubits are used to simulate such dynamics. For example, in \,\cite{perez} $3$ ancilla are used for a decohering qubit (under simultaneous damping under the 3 Pauli operators). Such an implementation needs $O(\log n)$ qubits for the system + $O(l)$ qubits for the bath. Below, we propose a scheme with reduction in number of environment qubits to $O(1)$ (i.e. constant) for any $l$.\\

Consider a Markovian quantum channel with Kraus representation,
\begin{equation}
    \Phi_t:\rho_S(t)=\sum_{i=1}^l\Omega_i\rho_S(0)\Omega_i^\dagger.
\end{equation}
The usual Stinespring dilation gives,
\begin{equation}
    U_{SE}=\sum_{i=1}^l\Omega_i\otimes\ket{i}\bra{0}_E.
\end{equation}
By implementing $U_{SE}$ and tracing out the environment, we can simulate the quantum channel. When we implement $U_{SE}$ gate, we are simultaneously implementing the Kraus operators. In \,\cite{transport}, a method is proposed where the simultaneous process is instead implemented sequentially using lesser number of qubits. The method is summarized below. In \,\cite{transport}, this method is used for simulating environment assisted quantum transport. Here, we have generalized it to any open quantum system.\\
Consider a channel given by,
\begin{equation}
    \rho_S(t)=\sum_{i=1}^l\Omega_i\rho_S(0)\Omega_i^\dagger.
\end{equation}
The Kraus representation can be decomposed as,
\begin{equation}\label{sequential}
    \Phi_t=\Phi_t^l\circ\Phi_t^{l-1}\circ\ldots\circ\Phi_t^1 
\end{equation}
where,
\begin{equation}
    \Phi_t^i(\rho_S)=\Omega_i\rho_S\Omega_i^\dagger + \Omega_i'\rho_S\Omega_i'^\dagger.
\end{equation}
Now,
\begin{equation}\label{compositionmap}
    \Phi_t^j\circ\Phi_t^i=\Omega_i\rho_S\Omega_i^\dagger+ \Phi_t^j(\Omega_i'\rho_S\Omega_i'^\dagger),
\end{equation}

for $\Omega_i'=\sqrt{I-\Omega_i^\dagger\Omega_i}$. And,
\begin{equation}
    \Omega_i'\approx I-\frac{1}{2}\Omega_i^\dagger\Omega_i.
\end{equation}
Thus,
\begin{equation}\label{approxomega'}
    \Omega_i'\rho_S\Omega_i'^\dagger\approx\rho_S-\frac{1}{2}\Omega_i^\dagger\Omega_i\rho_S-\frac{1}{2}\rho_S\Omega_i^\dagger\Omega_i.
\end{equation}
To prove \,\,Eq.\eqref{sequential}, consider the second term in \,\,Eq.\eqref{compositionmap},
\begin{equation}
    \Phi_t^j(\Omega_i'\rho_S\Omega_i'^\dagger)=\Omega_j(\Omega_i'\rho_S\Omega_i'^\dagger)\Omega_j^\dagger + \Omega_j'(\Omega_i'\rho_S\Omega_i'^\dagger)\Omega_j'^\dagger.
\end{equation}
Substituting \,\,Eq.\eqref{approxomega'} and taking terms only upto first order we get,
\begin{equation}
\begin{split}
    \Phi_t^j(\Omega_i'\rho_S\Omega_i'^\dagger)=\Omega_j\rho_S\Omega_j^\dagger+\rho_S&-\frac{1}{2}\Omega_i^\dagger\Omega_i\rho_S-\frac{1}{2}\rho_S\Omega_i^\dagger\Omega_i\\
    &-\frac{1}{2}\Omega_j^\dagger\Omega_j\rho_S-\frac{1}{2}\rho_S\Omega_j^\dagger\Omega_j.
\end{split}
\end{equation}
Substituting back in \,\,Eq.\eqref{compositionmap},
\begin{equation}
\begin{split}
    \Phi_t^j\circ\Phi_t^i=\Omega_i\rho_S\Omega_i^\dagger+\Omega_j\rho_S\Omega_j^\dagger+\rho_S\\
    -\frac{1}{2}\Omega_i^\dagger\Omega_i\rho_S-\frac{1}{2}\rho_S\Omega_i^\dagger\Omega_i\\
    -\frac{1}{2}\Omega_j^\dagger\Omega_j\rho_S-\frac{1}{2}\rho_S\Omega_j^\dagger\Omega_j.
\end{split}
\end{equation}
Composing all maps, Eq.(\ref{sequential}) will give,
\begin{equation}\label{proof}
    \begin{split}
        \Phi_t^l\circ\Phi_t^{l-1}\circ\ldots\circ\Phi_t^1=&\sum_{i=1}^l\Omega_i\rho_S\Omega_i^\dagger+\rho_S\\
        -\frac{1}{2}&\sum_{i=1}^l\Omega_i^\dagger\Omega_i\rho_S-\frac{1}{2}\sum_{i=1}^l\rho_S\Omega_i^\dagger\Omega_i.
    \end{split}
\end{equation}
Now, $\sum_{i=1}^l\Omega_i^\dagger\Omega_i=I$. Thus, the last two terms above are $=\frac{1}{2}\rho_S$. 

Sum of last three terms in \,\,Eq.\eqref{proof}: $\rho_S-\frac{1}{2}\rho_S-\frac{1}{2}\rho_S=0$.

Thus,
\begin{equation}
    \Phi_t^l\circ\Phi_t^{l-1}\circ\ldots\circ\Phi_t^1=\sum_{i=1}^l\Omega_i\rho_S\Omega_i^\dagger=\Phi_t.
\end{equation}
\begin{figure}
    \centering
    \includegraphics[width=0.5\textwidth]{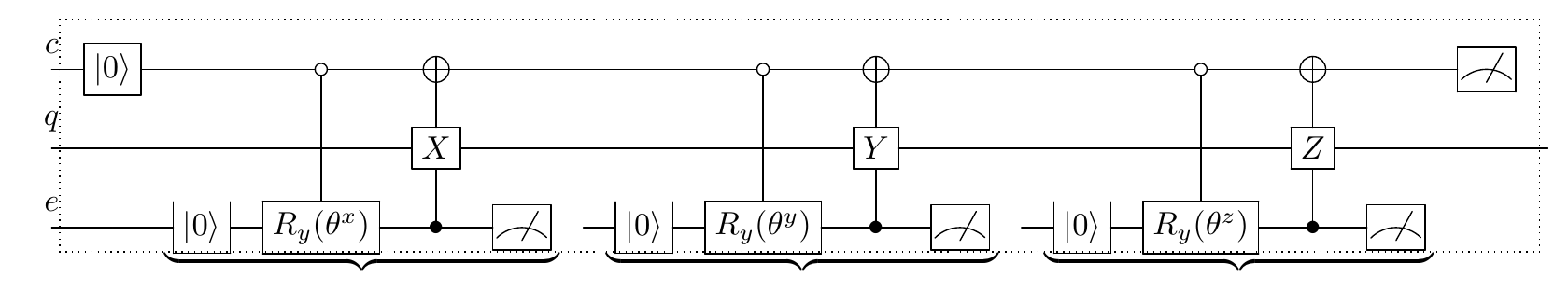}
    \caption{Quantum circuit for optimizing resource requirements (qubits and gates) by converting a parallel process in to a sequential one (\,\,Eq.\ref{sequential}). Here, decoherence of a qubit is modeled. $c$ is the control qubit, $q$ is the system, and $e$ simulates the environment. One step(box) consists of sequential implementation of appropriate $X$, $Y$, and $Z$ gates.}
    \label{fig:optimization}
\end{figure}
This proves that we can sequentially implement Kraus operators to simulate a quantum channel. Fig.\ref{fig:optimization} shows the quantum circuit for a decohering qubit. Here, we have sequentially applied the Kraus operations using an extra control qubit $c$. This helps in implementing the decomposition map so that $\ket{0}$ state of $c$ is used to simulate Kraus operators sequentially (using the composition of maps \,\,Eq.\eqref{sequential}). $e$ is traced out and reset after implementing each Pauli operator and $c$ is traced out after one complete step.  Here, we have shown the quantum circuit for Markovian evolution. The same can be used for Non-Markovian evolution, by adding environment qubits, as illustrated in the previous section.\\

{\it Complexity Analysis:}\\
We need $\log n$ qubits for implementing a quantum system with Hilbert space $dim=n$. We can implement the effect of environment by using $k$ qubits (as described in previous section), where $k$ is the order upto which we are taking memory effects. For optimization using sequential implementation, one extra qubit is needed as control. Thus,\\
Total number of qubits $=O(\log n + k)$\\

To implement each Kraus operation, we need $2$ gates (rotation of environment qubit + implementation of operation). If there are $l$ operators in the Kraus representation, we will need $O(l)$ gates. This holds for Markovian processes. For Non-Markovian evolution, we need additional $O(k)$ gates ($k$ rotation gates for memory, plus $k-1$ SWAP gates for updating the environment after each step, see Fig.\,\ref{fig:1step}). Thus,\\
Number of gates in one step= $O(k+l)$ \\
Number of gates in complete evolution= $O(T(k+l))$, where $T$ is the total (discrete-)time.\\

The quantum complexity obtained above is a reduction over the usual method for simulation. For example, $O(\log n + k\log l)$ qubits are needed to simulate a quantum channel using the existing method of Stinespring dilation. Further, Suzuki-Lie Trotter decomposition of general unitary gates over q number of qubits takes $O(\exp{q})$ gates. Thus, for Stinespring dilation $O(l^k)$ gates are needed. \\

Thus, the quantum circuits proposed in this article, along with a method for sequential implementation of operations can be used to simulate both Markovian and Non-Markovian open quantum systems. Further, the obtained circuits are optimal in both time, and space requirements.

%=================================================================================================================
\section{Conclusion}\label{sec:conclusion}
We have presented a general framework to develop quantum circuits for digital quantum simulation of Markovian and Non-Markovian open quantum systems. The environment is simulated using ancilla qubits which control the dynamics on the system. For Markovian dynamics, the environment is traced out after each step and reset to $\ket{0}$. For Non-Markovian systems, the environment is only partially traced out so as to retain information for the next step. This helps create a memory of the system, which is updated at the end of each step using SWAP gates. Numerical simulations of these models reproduce the characteristic monotonic behavior for Markovian systems, and oscillations due to memory effects in Non-Markovian systems. Further, we have introduced a method to optimize resource requirements for simulating quantum channels. The method relies of converting a parallel process to a sequential one, thus reducing space requirements. It employs simple quantum gates (controlled rotation and CNOTs) which reduce time complexity when compared to the widely used Stinespring dilation method. The Markovian and Non-Markovian quantum circuits, together with a method for optimizing complex processes provides an optimal framework to simulate any open quantum system.\\

Open quantum systems are generally simulated on analog quantum simulators, with digital quantum simulation being studied only recently. The models presented in this article provide a promising framework to simulate open quantum systems on digital-NISQ technology. For example, due to small resource requirement, the quantum circuits can be used to implement complex open quantum systems on IBM Q-Experience\,\cite{perez}. The framework can also be used to understand Non-Markovian dynamics in various contexts, by tuning parameters like $\theta^i$, $k$, etc., since it is simpler to use than the more common HEOM (Hierarchical equations of motion), Nakajima-Zwanzig equation, or other models. An interesting line of work could be to explore how the different theoretical models relate to each other. Further, it might be useful to study the implications of the optimization presented here, to larger scale quantum computing. Another potential application could be to study reservoir engineering on quantum simulators, so that the memory effects can be used to create long-term coherence as
 observed in Fig.\,\ref{fig:interesting}.

\end{document}